\documentclass{elsart}
\journal{Physics Letters B}

\usepackage{natbib,amssymb,graphicx}

\begin{document}

\begin{frontmatter}
\title{Hadron formation and attenuation in deep inelastic lepton scattering 
off nuclei}
\author{T. Falter\corauthref{cor}},
\corauth[cor]{Corresponding author.}
\ead{Thomas.Falter@theo.physik.uni-giessen.de}
\author{W. Cassing},
\author{K. Gallmeister},
\author{U. Mosel}
\address{Institut fuer Theoretische Physik, Universitaet Giessen, D-35392 Giessen, Germany}

\begin{abstract}
We investigate hadron formation in deep inelastic
lepton scattering on $N$, $Kr$ and $Xe$ nuclei in the kinematic regime
of the HERMES experiment. The elementary electron-nucleon
interaction is described within the event generator PYTHIA while a
full coupled-channel treatment of the final state interactions is
included by means of a BUU transport model. We
find a good agreement with the measured charged hadron multiplicity ratio
$R_M^h$ for $N$ and $Kr$ targets by accounting for the deceleration and 
absorption of the primarily produced particles as well as for the creation 
of secondary hadrons in the final state interactions. 
\end{abstract}
\begin{keyword}
hadron formation \sep hadron attenuation \sep deep inelastic scattering 
\sep electroproduction \sep nuclear reactions \sep meson production
\PACS 24.10.-i \sep 25.30.-c \sep 13.60.Le\\
\end{keyword}
\end{frontmatter}

Hadron production in deep inelastic lepton-nucleus
scattering (DIS) offers a promising tool to study the physics of
hadronization \cite{Kop}. The reaction of the exchanged virtual photon
(energy $\nu$, virtuality $Q^2$) with a bound nucleon leads to the
production of several hadrons. While the primary production is determined
by the fragmentation function -- in medium possibly different from that
in vacuo -- the number of ultimately observed hadrons and their energy 
distribution depends also on their rescattering in the surrounding nuclear
medium. Consequently, the particle spectrum of a lepton-nucleus interaction 
will differ from that of a reaction on a free nucleon. In order to
explore such attenuation effects the HERMES collaboration has
investigated the energy $\nu$ and fractional energy $z_h=E_h/\nu$
dependence of the charged hadron multiplicity ratio
\begin{equation}
\label{eq:multiplicity-ratio}
R_M^h(z,\nu)=\left(\frac{N_h(z,\nu)}{N_e(\nu)}\right)_A\bigg/\left(\frac{N_h(z,\nu)}{N_e(\nu)}\right)_D
\end{equation}
in DIS off $N$ \cite{HERMES1} and $Kr$ \cite{HERMES2} nuclei. Here $N_h(z,\nu)$
represents the number of semi-inclusive hadrons in a given ($z,\nu$)-bin and
$N_e(\nu)$ the number of deep inelastically scattered leptons in the same 
$\nu$-bin. It was suggested in Ref. \cite{HERMES1}, that a phenomenological
description of the $R_M^h$ data can be achieved if the formation
time, i.e. the time that elapses from the moment when the photon
strikes the nucleon until the reaction products have evolved to
physical hadrons, is assumed to be proportional to $(1-z_h)\nu$ in
the target rest frame. This $(1-z_h)\nu$ dependence of the formation time 
$\tau_f$ is compatible with the gluon-bremsstrahlung model of Ref. \cite{Kop}.
In the investigations of Ref. \cite{HERMES1} any interaction of the reaction 
products with the remaining nucleus during this formation time has been 
neglected. After the formation time the hadrons could get absorbed according 
to their full hadronic cross section. Another interpretation
of the observed $R_M^h$ spectra -- as being due to a combined effect of a 
rescaling of the quark fragmentation function in nuclei due to partial 
deconfinement as well as the absorption of the produced hadrons -- has
recently been given by the authors of Ref. \cite{Acc02}. Furthermore,
calculations based on a pQCD parton model \cite{Wang,Arl03} explain the 
attenuation observed in the multiplicity ratio solely by partonic multiple 
scattering and induced gluon radiation completely neglecting any hadronic 
final state interactions (FSI). It has already been pointed out by 
the authors of Ref. \cite{Acc02} that a shortcoming of the existing models is 
the purely absorptive treatment of the FSI. We 
will avoid this problem by using a semi-classical coupled-channel transport
model \cite{Eff99} which already has been employed to describe
high energy photo- \cite{Photo} and electroproduction \cite{Fal02}
off nuclei.

We point out that the formation time also plays an important 
role in studies of ultra-relativistic heavy-ion reactions. For example, the 
observed quenching of high transverse momentum hadrons in $Au+Au$ reactions 
relative to $p+p$ collisions is often thought to be due to jet quenching in 
a quark gluon plasma. However, the attenuation of high $p_T$ hadrons might 
at least partly be due to hadronic rescattering processes \cite{Gal03,Cas03}.

Unfortunately, the average hadron formation time is not well known
and the number of $\tau_f \approx1$ fm/c as used commonly in the
Bjorken estimate for the energy density \cite{bjorken} is nothing
but an educated guess. The nonperturbative nature of this number
-- due to time scales of $\sim $ 1 fm/c and hadronic size scales
of 0.5--1 fm -- excludes perturbative evaluation schemes; it is
hard to calculate $\tau_f$ from first principles and formation
times  cannot be addressed in present lattice QCD simulations. One
might expect that the rather successful string models \cite{LUND}
shed some further light on this number, since the intrinsic time
scale $\tau_0$ for the $q-\bar{q}$ formation vertex can be related
to the fragmentation function and string tension, respectively
\cite{Anders}. However, the actual parameters employed in
current transport codes are not unique, with hadron formation times 
ranging from 0.3--2 fm/c \cite{Cas99,UrQMD1}, depending on the flavor,
momentum and energy of the created hadrons. In fact, the rapidity 
and transverse mass spectra from relativistic nucleus-nucleus collisions 
are not very sensitive to the formation time $\tau_f$ \cite{Cas03}. It is 
therefore essential to check whether these times are compatible with 
constraints extracted from reactions, where the collision geometry is much 
better under control. 

To this aim the attenuation of antiprotons produced in $p+A$ reactions at 
the AGS energies of 12.3 GeV and 17.5 GeV has been investigated on various
nuclear targets in Ref. \cite{Cas02} and a range of values for
$\tau_f= 0.4 - 0.8$ fm/c has been extracted in comparison to the
data from the E910 Collaboration \cite{E910}.

In our present approach the lepton-nucleus interaction is split
into two parts: 1) In the first step the virtual photon is absorbed
on a bound nucleon of the target; this interaction produces a
bunch of particles that in step 2) are propagated within the
transport model. Coherence length effects in the entrance channel,
that give rise to nuclear shadowing, are taken into account as
described in Ref. \cite{Fal02}. The virtual photon-nucleon
interaction itself is simulated by the Monte Carlo generator
PYTHIA v6.2 \cite{pythia} which well reproduces the experimental
data of Ref. \cite{HERMES3} on a hydrogen target. Depending on
whether the photon interacts directly with the nucleon or via one
of its hadronic fluctuations ($\rho^0, \omega, \Phi, J/\Psi$ or a
perturbative $q\overline{q}$ fluctuation) the 
reaction leads to the excitation of one or more hadronic strings 
which in our approach are assumed to fragment very rapidly into colorless 
prehadrons. 

The time that the reaction products need to evolve
to physical hadrons, i.e., the production time $t_p$ of the prehadrons 
plus the time needed to build up the hadronic wave function, we 
denote as formation time $t_f$ in line with the
convention in transport models. For simplicity we assume that the
formation time is a constant $\tau_f$ in the rest frame of each
hadron and that it does not depend on the particle species. 
Due to time dilatation the formation
time $t_f$ in the laboratory frame is then proportional to the
particle's energy
\begin{equation}
\label{eq:formation-time}
        t_f=\gamma\cdot\tau_f=\frac{z_h\nu}{m_h}\cdot\tau_f .
\end{equation}
The size of $\tau_f$ can be estimated by the time that the constituents 
of the hadrons need to travel a distance of a typical hadronic radius
(0.5--0.8 fm). 

\begin{figure}
  \begin{center}
       \includegraphics[width=10cm]{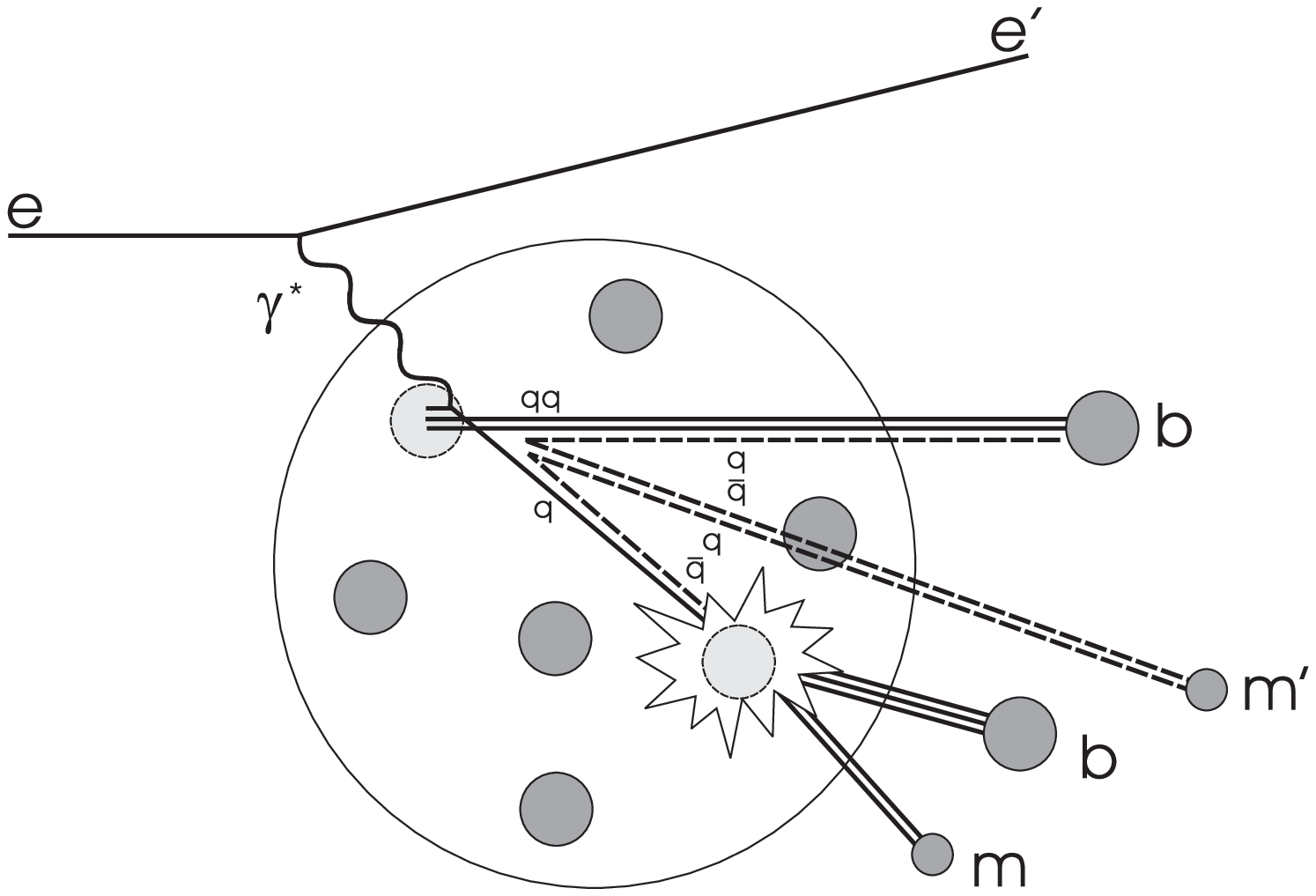}
  \end{center}
  \caption{Illustration of an electron nucleus interaction: The virtual 
photon $\gamma^*$ excites a hadronic string by hitting a quark $q$ inside a 
bound nucleon. In our example the string between the struck quark $q$ and 
diquark $qq$ fragments due to the creation of two quark-antiquark pairs.
One of the antiquarks combines with the struck quark to form a 'leading' 
pre-meson $m$, one of the created quarks combines with the diquark to form a 
'leading' pre-baryon $b$. The remaining partons combine to a pre-meson $m'$ 
that, depending on the mass of the meson, might leave the nucleus before
it hadronizes (see Eq. (\ref{eq:formation-time})). Note that in our approach 
the actual production time $t_p$ of the non-leading prehadrons has no 
effect on our results since we neglect any interaction until $t_f$. See text
for details.} 
  \label{fig:fig1}
\end{figure}

To illustrate the situation shortly after the photon nucleon reaction
Fig.~\ref{fig:fig1} shows the excitation and fragmentation of a hadronic 
string in a deep inelastic scattering process. For simplicity we do not
show any gluon bremsstrahlung of the struck quark in this figure. Note, 
however, that the possibility of such final state gluon radiation plus 
subsequent $q\bar{q}$ splitting is included in the PYTHIA part of our model 
and leads to the creation of additional strings
\cite{pythia}. It has been emphasized in Ref. \cite{Cio02} that the string 
propagating through the nucleus is a rather short (white) object of length
$\approx 1$ fm since the slow end of the string is accelerated very fast. 
When the primary string fragments -- due to the creation of $q\bar{q}$ pairs 
from the vacuum -- new colorless prehadrons are produced, which we propagate
in space-time.

As discussed in Ref. \cite{Kop03} the production time 
$t_p$ of these prehadrons has to be distinguished from the total formation 
time $t_f$ of the final hadrons which is dilated according to Eq. 
(\ref{eq:formation-time}). It has been confirmed \cite{Bia87} within the Lund 
model that the production time $t_p$ vanishes for $z_h\to 0$ and 
$z_h\to 1$. In the present numerical realization of our model we first 
approximate this behavior by setting the production time $t_p$ to zero for 
{\it all} prehadrons, but  will also discuss the effect of a finite production 
time at the end of this study.

Right after the photon nucleon interaction the primary string
should interact with a hadronic cross section because its transverse 
size is essentially that of the original nucleon. Motivated by the 
constituent quark model we assume that this hadronic cross 
section is shared by the quark/diquark at the string ends and after the 
fragmentation by the so called leading prehadrons that contain this 
quark or diquark. Our PYTHIA simulations show that in most cases the 
prehadrons with $z_h\approx 1$ are such leading hadrons since they
contain constituents (valence- or sea-quarks) from the target nucleon or 
the hadronic component of the photon. They can therefore interact directly 
after the photon-nucleon interaction with a constant effective cross section 
which we denote as $\sigma_\mathrm{lead}$. The cross sections of the other 
prehadrons, 
that solely contain quarks and antiquarks created from the vacuum, emerge at 
intermediate $z_h$. They are assumed to be non-interacting until $t_f$. This 
assures that the summed cross section of 
the complete final state right after the photon-nucleon interaction is 
approximately that of the original nucleon. Each time when a new hadron
has formed, the summed cross section rises just like in the approach 
of Ref. \cite{Cio02}. After the hadron formation time $t_f$ by definition 
all hadrons interact with their full hadronic cross section $\sigma_h$. Note,
that our concept of leading hadrons is in accordance with those of other 
transport models for high energy reactions \cite{Cas03,Cas99,Geiss,URQMD2}. 

Since the lighter (intermediate $z_h$) hadrons have large formation times in
the target frame (see Eq. (\ref{eq:formation-time})) they may escape the 
nucleus without being attenuated if they are non-leading. However, many 
($\approx 2/3$) of the observed hadrons with intermediate $z_h$ are not 
directly produced in the string fragmentation but stem from decays of the 
much heavier vector mesons with
correspondingly shorter formation times. These vector mesons may
therefore form inside the nuclear volume and thus be subjected to FSI.
The effect of the FSI, finally, will
depend dominantly on the nuclear geometry, i.e. the size of the
target nucleus.

In our present study the FSI are described by a coupled-channel
transport model based on the Boltzmann-Uehling-Uhlenbeck (BUU)
equation which describes the time evolution of the phase space density 
$f_i(\vec r,\vec p,t)$ of particles of type $i$ that can interact via binary 
reactions. These particles involve nucleons, baryonic 
resonances and mesons ($\pi$, $\eta$, $\rho$, $K$, ...) that are produced 
either in the primary reaction or during the FSI. In this work we also 
have to account for the prehadrons emerging from the string fragmentation. 
For a particle species $i$ the BUU equation takes the form
\begin{equation}
  \left(\frac{\partial}{\partial t}+\frac{\partial H}{\partial\vec p}\frac{\partial}{\partial \vec r}-\frac{\partial H}{\partial \vec r}\frac{\partial}{\partial \vec p}\right)f_i(\vec r,\vec p,t)=I_{coll}[f_1,...f_i,...,f_M],
\end{equation}
where the Hamilton function $H$ includes a position and momentum dependent 
mean-field potential for baryons. The collision integral on the 
right hand side accounts for the creation and annihilation of particles of 
type $i$ in a collision as well as elastic scattering. The transition rates 
are determined from the particular (vacuum) cross sections. For fermions 
Pauli blocking is taken into account in $I_{coll}$ via blocking factors. The
prehadrons are treated like ordinary hadrons except for their modified 
interaction cross section and the fact that they are not allowed to decay 
during the formation time. The BUU equations of each particle species $i$ are 
coupled via the collision integral and the mean field in case of baryons. The 
resulting system of coupled differential-integral equations is solved via a 
test particle ansatz for the phase space density. For further details of the 
transport
model we refer the reader to Ref.~\cite{Eff99}. 

Most of the hadronic FSI 
happen with invariant energies $\sqrt{s} \geq 2.2$ GeV and are described 
within the Lund string formation and decay scheme \cite{LUND} as also 
implemented in the transport approaches \cite{Eff99,Photo,Fal02} as well
as \cite{Cas99,Cas02,Geiss}. 
The important difference between a purely absorptive treatment of the FSI and
the coupled-channel description provided by the BUU model is that in
an interaction with a nucleon a hadron might not only be absorbed or recreated
but also be decelerated in an elastic or inelastic collision. Furthermore, it
may in addition produce several low energy particles. In the case
of electroproduction of hadrons these interactions leads to a redistribution
of strength from the high $z_h$ part of the hadron energy spectrum
to lower values of the energy fraction $z_h$.

In our calculation we employ all kinematic cuts of the HERMES
experiment as well as the geometrical cuts of the detector. In
actual numbers:  we require for the Bjorken scaling variable
$x=\frac{Q^2}{2m_N\nu}>0.06$ (with $m_N$ denoting the nucleon
mass), for the photon virtuality $Q^2>1$ GeV$^2$ and for the
energy fraction of the virtual photon $y=\nu/E_{beam}<0.85$. In
addition, the PYTHIA model introduces a lower cut in the invariant
mass of the photon-nucleon system at $W=4$ GeV that is above the
experimental constraint $W>2$ GeV. This limits our calculations to
minimal photon energies of $\nu_{min}=8.6$ GeV as compared to
$\nu_{min}=7$ GeV in the HERMES experiment and leads to a
suppression of high $Q^2$ events at energies below
$\nu\approx 15$ GeV. 

\begin{figure}
  \begin{center}
    \includegraphics[width=12cm]{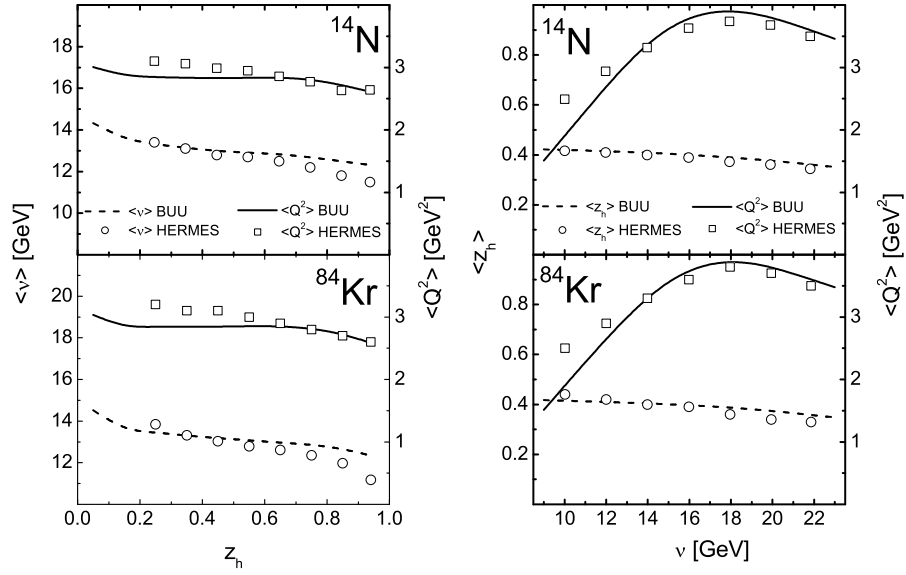}
  \end{center}
  \caption{Model predictions for the average values of the kinematic variables
in charged hadron production in comparison with the experimental numbers at 
HERMES. For the calculation we used the formation time $\tau_f=0.5$ fm/c and 
a leading prehadron cross section $\sigma_\mathrm{lead}=0.33\sigma_h$. 
{\it Left:} $\langle\nu\rangle$ and $\langle Q^2\rangle$ as a function 
of $z_h$ compared to the experimental values for $N$ and $Kr$ targets
 \cite{private}. 
{\it Right:} Same for $\langle z_h\rangle$ and $\langle Q^2\rangle$ as a 
function of $\nu$.}
  \label{fig:fig2}
\end{figure}

In Fig. \ref{fig:fig2} we present the average values of $Q^2$ and 
$\nu$ ($z_h$) for our simulated event samples on $N$ and $Kr$ as a function 
of $z_h$ ($\nu$) in comparison to the experimental quantities \cite{private}.
In order to compare with the $\nu$ dependence of $\langle Q^2\rangle$ 
and $\langle z_h\rangle$ we only account for
hadrons with $z_h>0.2$ as in the HERMES experiment. For both the
$N$ and the $Kr$ target the average kinematical variables are well reproduced 
by our model. The underestimation 
of $\langle Q^2\rangle$ at low photon energies $\nu<12$ GeV is due to the
PYTHIA cut in $W \geq 4$ GeV which suppresses higher values of
$Q^2$ at low photon energies $\nu$. This is also the reason why 
the average value of $Q^2$ in the $z_h$ spectrum comes out slightly too 
low within our model.

\begin{figure}
  \begin{center}
    \includegraphics[width=10cm]{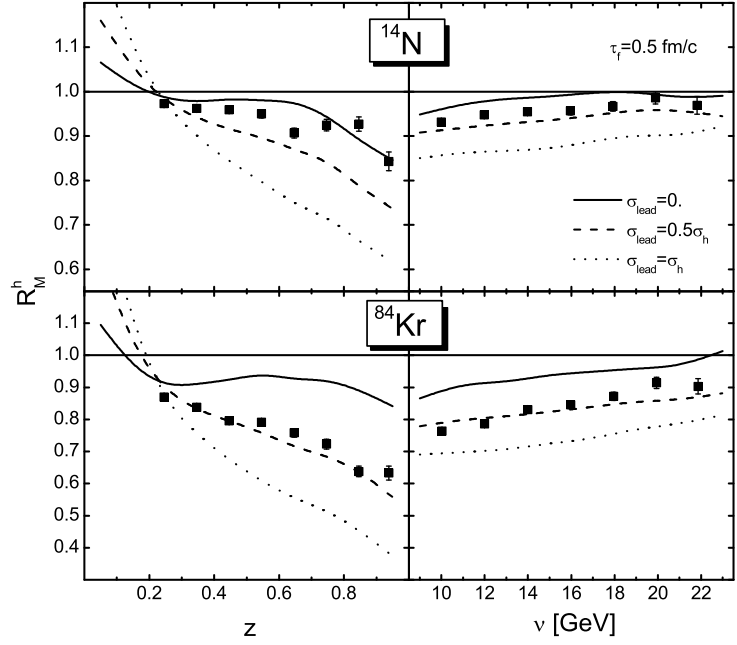}
  \end{center}
  \caption{Calculated multiplicity ratios (Eq. (\ref{eq:multiplicity-ratio})) of 
charged hadrons for $N$ and $Kr$ targets for 
fixed formation time $\tau_f=0.5$ fm/c and different values of the leading 
prehadron cross section:
$\sigma_\mathrm{lead}=0.$, i.e. without any prehadronic interaction (solid line),  
$\sigma_\mathrm{lead}=0.5\sigma_h$ (dashed line) and  $\sigma_\mathrm{lead}=\sigma_h$ 
(dotted line). The data for the Nitrogen target have been taken from 
Ref. \protect\cite{HERMES1} while the Krypton data stem from 
Ref. \protect\cite{HERMES2}.}
  \label{fig:fig3}
\end{figure}

\begin{figure}
  \begin{center}
    \includegraphics[width=10cm]{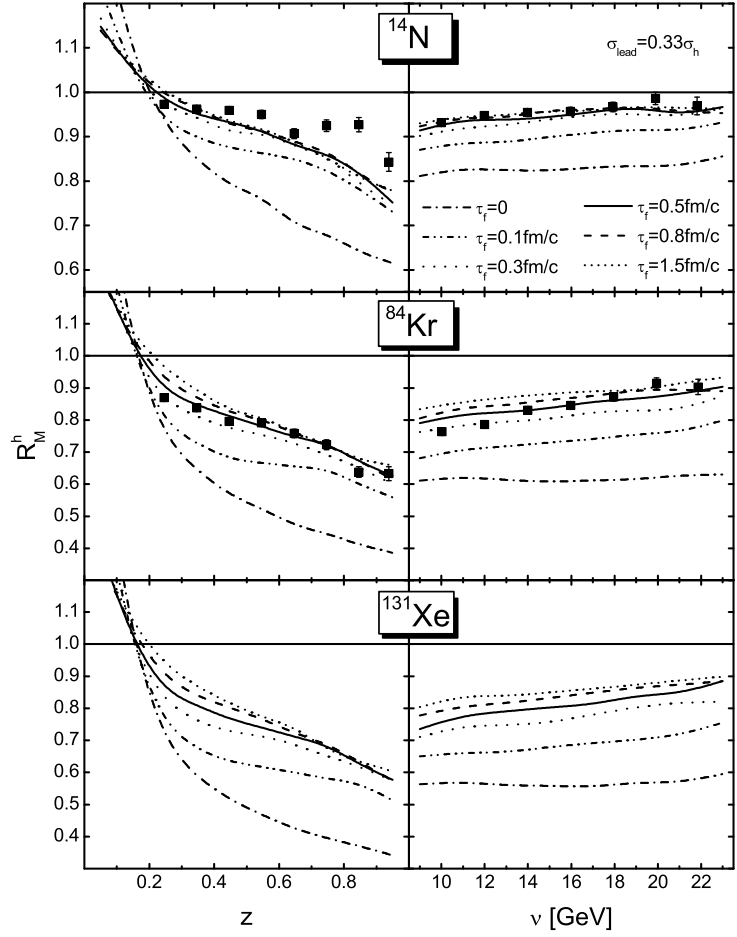}
  \end{center}
  \caption{Calculated multiplicity ratios of charged hadrons for $N$, $Kr$ and $Xe$ 
targets for a fixed leading prehadron cross section 
$\sigma_\mathrm{lead}=0.33\sigma_h$ and different values of the formation time from $\tau_f=0$ to 1.5 fm/c. 
The data are the same as in Fig. \ref{fig:fig3}.}
\label{fig:fig4}
\end{figure}

We proceed with a discussion of the actual results of our PYTHIA + BUU
simulations. In Fig. \ref{fig:fig3} we show the calculated
multiplicity ratio $R_M^h$ for $N$ and $Kr$ targets using an
'estimated' formation time of 0.5 fm/c and different values for
the leading prehadron cross section $\sigma_\mathrm{lead}$. The data have been
taken from Refs. \cite{HERMES1,HERMES2}.
Since the particles with $z_h$ close to 1 are predominantly leading hadrons
we can use the high $z_h$ part in the fractional energy spectrum
to obtain information on $\sigma_\mathrm{lead}$. 
The data for both nuclei indicate that $\sigma_\mathrm{lead}$ has to be in 
the range
0--0.5 $\sigma_h$ with $\sigma_h$ ($h=\pi^\pm,K^\pm,p,\ldots$) taken from 
\cite{PDG}. For the heavier 
nucleus, $Kr$, we clearly underestimate the hadron attenuation with 
$\sigma_\mathrm{lead}$=0
since most of the particles, especially those with large energies, escape the 
nucleus due to time dilatation. If one wants to describe the strong 
attenuation of hadrons at large fractional energy $z_h$ without any 
prehadronic interactions one would need an unphysical short 
formation time $\tau_f<0.1$ fm/c. This, however, is ruled out 
by the measured $\nu$ dependence of $R_M^h$ since a vanishing formation time 
leads to a multiplicity ratio $R_M^h$ which is considerably too low (see 
dash-dotted and dash-dot-dotted curves in Fig. \ref{fig:fig4}).

The dotted line in Fig. \ref{fig:fig3} shows the result of a calculation 
where the leading prehadrons interact with the full hadronic cross section 
$\sigma_h$; this leads to a too strong attenuation of charged
hadrons. A good agreement with the data is 
achieved for $\sigma_\mathrm{lead}=0.33\sigma_h$ during the formation time
$\tau_f$ as can be seen in Fig. \ref{fig:fig4}. We note, that this value for
$\sigma_\mathrm{lead}$ represents an average value over time from the virtual 
photon-nucleon interaction to the actual hadron formation time. For a 
detailed investigation we refer the reader to a forthcoming study \cite{F2}.

In Fig. \ref{fig:fig4} we investigate the influence of different formation
times $\tau_f$ on $R_M^h$ using the effective cross section 
$\sigma_\mathrm{lead}=0.33\sigma_h$. We find, that formation times 
$\tau_f\gtrsim 0.3$ fm/c are needed to describe the experimental data with 
little sensitivity to higher values. This is compatible to the range of 
values extracted from the antiproton attenuation studies in Ref. 
\cite{Cas02}. The steep rise of $R_M^h$ for $z_h\lesssim 0.2$ is caused by 
the energy loss and the production of low energy secondary particles in 
elastic and inelastic FSI. We mention that in models dealing with purely
absorptive FSI a ratio much larger than one can only be explained by a 
drastic change of the fragmentation function.

We note that the photon energy dependence 
of the ratio $R_M^h$ for the $Kr$ target is less well reproduced for energies 
below 14 GeV. The reason for this can be traced back 
to the $W \geq 4$ GeV cut of the PYTHIA model. Our simulations show that
the number of leading hadrons decreases with $Q^2$ for fixed
energy, but the W$\geq 4$ GeV cut discards larger values of $Q^2$
at energies $\nu<15$ GeV. For higher values of $Q^2$ the
importance of DIS events rises as compared to events, where the
photon interacts via a vector meson fluctuation (VMD events). In
the latter case one has initially 5 constituent quarks and antiquarks and 
thus gets more leading prehadrons than in a DIS event. This leads to a 
mismatch of leading and nonleading prehadrons in the region 
$z_h\lesssim 0.4$. Since the number of prehadrons created in an 
electron-nucleon collision decreases exponentially with increasing $z_h$, 
this region contributes dominantly to the $z_h$ integrated $\nu$ spectrum. 
This deficiency of the present model will be cured in a more detailed 
upcoming work \cite{F2}.

In the pQCD parton model of Ref. \cite{Wang} the multiple parton 
scattering leads to a modification of the fragmentation function and
predicts a hadron attenuation $\sim A^{2/3}$. In our approach, however, 
the fragmentation is assumed to be decided on time scales of the nucleon 
dimension itself such that only the 'free' fragmentation function enters. 
All attenuation effects then are attributed to FSI of the leading and
secondary (pre-)hadrons. In order to distinguish experimentally between the 
different concepts, it is thus important to get the scaling with target mass 
$A$. To this aim Fig. \ref{fig:fig4} also shows predictions for a $Xe$ target.
In accordance with the authors of Ref. \cite{Acc02} we predict only a small 
change in the multiplicity spectra compared to the $Kr$ target such that the 
scaling exponent is lower than $2/3$.

\begin{figure}
  \begin{center}
    \includegraphics[width=10cm]{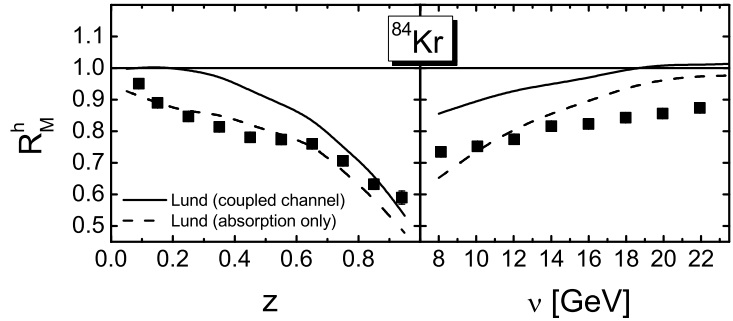}
  \end{center}
  \caption{Calculated multiplicity ratios of charged hadrons for a $Kr$ target 
assuming the finite production time $t_p$ given by Eq. (\ref{eq:lund-time}). 
For $t\geq t_p$ the prehadrons interact 
  with their full hadronic cross section. The dashed line shows the result of a simulation
   with a purely absorptive treatment of the FSI. The solid line represents the coupled-channel result.}
\label{fig:fig5}
\end{figure}

Finally, we discuss the effect of a finite production time $t_p$ (in the lab 
frame) for prehadrons
which we adopt, for consistency, from the Lund model \cite{Bia87}:
\begin{equation}
\label{eq:lund-time}
        t_p=\left(\frac{\ln(1/z^2)-1+z^2}{1-z^2}\right)\frac{z\nu}{\kappa}.
\end{equation}
Here $\kappa\approx 1$ GeVfm$^{-1}$ denotes the string tension. In the 
following 
we assume no interaction before $t_p$ and the full hadronic cross section for
prehadrons for $t\geq t_p$, i.e., there is no dependence 
on the formation time $t_f$ anymore. The dashed line in Fig. \ref{fig:fig5} 
represents the result of a Glauber-like treatment of the FSI where every time 
a prehadron interacts with another particle it 
is removed from the outgoing channel. As the authors of 
Ref. \cite{Acc02} we get a good description of the $z$-dependence of the 
multiplicity ratio. However, the $\nu$ spectrum is not attenuated strongly 
enough at the higher $\nu$ (r.h.s. of Fig. \ref{fig:fig5}). The solid curves 
show the effect of the coupled channels, i.e.,  
a particle is not only absorbed in a collision but produces a bunch of low 
energy particles, thereby shifting strength to the low $z$ part of the 
spectrum and thereby underestimating the attenuation at low $z$. Similarly, 
the attenuation is also too weak in the $\nu$-spectrum. Since an additional 
formation time with reduced cross sections would further enhance these 
discrepancies, we conclude that the data cannot be described with the 
production time (\ref{eq:lund-time}). For a detailed 
investigation of finite production times and alternative models we refer the 
reader to a forthcoming work \cite{F2}.

In summary, we have shown that one can describe the experimental data of the 
HERMES collaboration for hadron attenuation on nuclei without invoking any 
changes in the fragmentation function due to gluon radiation. 
In our dynamical studies, that include the most relevant FSI, we employ only 
the 'free' fragmentation function on a nucleon and attribute the hadron 
attenuation to the deceleration of the produced (pre-)hadrons due to FSI in 
the surrounding medium. We find that in particular the $z$-dependence of 
$R_M^h$ is very sensitive to the interaction cross section of leading 
prehadrons and can be used to determine $\sigma_\mathrm{lead}$. The 
interaction of the leading prehadrons during the formation time could be 
interpreted as an in-medium change of the fragmentation function, which
however could not be given in a closed form. The extracted average hadron 
formation times 
of $\tau_f \gtrsim 0.3$ fm/c are compatible with the analysis of antiproton 
attenuation in $p+A$ reactions at AGS energies \cite{Cas02}. In an upcoming 
work we will investigate in detail the spectra for different particle
species ($\pi^{\pm},K^{\pm},p,\overline{p}$) to examine, if the formation 
times of mesons and antibaryons are about equal. In addition we will improve 
our model to describe the primary 
photon-nucleon reaction below the PYTHIA threshold of $W\geq$ 4 GeV.

The authors acknowledge valuable discussions with N. Bianchi, E. Garutti,
C. Greiner, V. Muccifora and G. Van Steenhoven. This work was supported by 
DFG.

\newpage

\end{document}